# Smart Material Implication Using Spin-Transfer Torque Magnetic Tunnel Junctions for Logic-in-Memory Computing

Raffaele De Rose[1,*], Tommaso Zanotti[2], Francesco Maria Puglisi[2], Felice Crupi[1], Paolo Pavan[2], Marco Lanuzza[1]

[1]DIMES, University of Calabria, Rende 87036, Italy
[2]DIEF, University of Modena and Reggio Emilia, Modena 41125, Italy
[*]Corresponding author. E-mail address: r.derose@dimes.unical.it

**Abstract**

Smart material implication (SIMPLY) logic has been recently proposed for the design of energy-efficient Logic-in-Memory (LIM) architectures based on non-volatile resistive memory devices. The SIMPLY logic is enabled by adding a comparator to the conventional IMPLY scheme. This allows performing a preliminary READ operation and hence the SET operation only in the case it is actually required. This work explores the SIMPLY logic scheme using nanoscale spin-transfer torque magnetic tunnel junction (STT-MTJ) devices. The performance of the STT-MTJ based SIMPLY architecture is analyzed by varying the load resistor and applied voltages to implement both READ and SET operations, while also investigating the effect of temperature on circuit operation. Obtained results show an existing tradeoff between error rate and energy consumption, which can be effectively managed by properly setting the values of load resistor and applied voltages. In addition, our analysis proves that tracking the temperature dependence of the MTJ properties through a proportional to absolute temperature (PTAT) reference voltage at the input of the comparator is beneficial to mitigate the reliability degradation under temperature variations.

## 1. Introduction

Material implication (IMPLY) logic is widely recognized as promising to realize fast energy-efficient Logic-in-Memory (LIM) architectures, while avoiding the von Neumann bottleneck related to data transfer from/to an off-chip non-volatile memory as occurs in conventional computing platforms [1–8]. Moreover, IMPLY and FALSE operations (i.e., the primitives of material implication logic) form a computationally complete logic basis [8]. This allows performing all logic computations by simply iterating these two basic operations, i.e., through multi-cycle operations. Nonetheless, the conventional IMPLY logic scheme relying on non-volatile resistive memory devices is hindered by severe issues, such as the degradation of the logic states and the very narrow design space related to $V_{SET}$ and $V_{COND}$ voltages to be used for proper operation [9]. To deal with above issues, an alternative smart implication (SIMPLY) logic scheme was recently proposed [9–11], which mainly consists of adding a comparator to the classical IMPLY architecture. Such a comparator is used to implement a preliminary READ operation, whose outcome is exploited to perform the subsequent SET operation only in the case it is actually required (i.e., when both inputs are '0'). As compared to the conventional IMPLY scheme, this allows strongly alleviating the issue of logic state degradation, breaking the tradeoff between the choice of applied $V_{SET}$ and $V_{COND}$ voltages, and saving energy at the cost of minimum penalty in terms of circuit area and complexity [9].

Although SIMPLY logic was originally proposed and validated for Resistive Random Access Memory (RRAM) devices [9–11], its feasibility using spin-transfer torque magnetic tunnel junction (STT-MTJ) devices was preliminarily investigated in our previous work [12]. As a result of our previous study, we demonstrated that the STT-MTJ based SIMPLY scheme enables faster, more energy-efficient and reliable logic operation as compared to its IMPLY counterpart. This work significantly extends the analysis presented in [12]. On the one hand, the performance of the STT-MTJ based SIMPLY architecture is analyzed by varying the load resistor ($R_G$) and applied voltages to implement both READ and SET operations (namely $V_{READ}$ and $V_{SET}$). On the other hand, the effect of temperature variations is also investigated for a given $R_G$ value and bias condition. As in [12], our analysis has been performed using a macrospin-based Verilog-A compact model [13] to describe the characteristics of a 30-nm perpendicular STT-MTJ device, including its temperature-dependent parameters.

The rest of the paper is structured as follows. Section 2 briefly describes the characteristics of the considered STT-MTJ device. Section 3 details the scheme and operation of the SIMPLY logic. Then, Section 4 reports and discusses the results of our simulation study. Finally, the main conclusions of the work are drawn in Section 5.

## 2. STT-MTJ Modeling

The sketch of a perpendicular STT-MTJ device with circular geometry is shown in Fig. 1. Its structure basically consists of three layers, i.e., a thin MgO oxide barrier sandwiched between two CoFeB ferromagnetic (FM) layers. One FM layer, namely reference layer (RL) has fixed magnetization orientation, whereas the other FM layer, namely free layer (FL) has variable magnetization orientation. This entails two stable configurations as determined by the relative magnetization orientation of the FL with respect to that of the RL (parallel or antiparallel). These two stable configurations involve two different resistance states, i.e., low resistance ($R_L$) in parallel state and high resistance ($R_H$) in antiparallel state (here corresponding to bit 1 and bit 0, respectively), whose difference is quantified by the tunnel magnetoresistance (*TMR*) ratio, as shown in Fig. 1. The switching from one state to the opposite is achieved through the STT mechanism, i.e., by applying a proper current flowing through the device [13].

To describe the STT-MTJ behavior and characteristics, we employed an analytical macrospin-based Verilog-A compact model [13]. Table 1 summarizes the main physical parameters of the 30-nm STT-MTJ device considered in this work [14–16]. As reported in Table 1, the model also accounts for the temperature dependence of physics parameters such as spin polarization factor (*P*), saturation magnetization ($M_S$), and interfacial perpendicular anisotropy constant ($K_i$). Such temperature-dependent parameters are modeled through semi-empirical laws [17, 18], as detailed in [19]. As a result of our modeling, Fig. 2(a)-(d) shows the trend of resistance and switching characteristics across temperatures. More specifically, Fig. 2(a) shows the temperature behavior of resistance values and corresponding *TMR* ratio at zero bias voltage. From this figure, $R_L$ is almost independent of temperature, whereas $R_H$ decreases with increasing temperature, as experimentally evidenced in [20–23]. This implies a *TMR* degradation at higher temperature, i.e., from 166% at 250 K down to 134% at 350 K as shown in Fig. 2(a). Fig. 2(b) and (c) show the behavior of the thermal stability factor ($\Delta$) and critical switching current ($I_c$), respectively. Within our macrospin modeling, such parameters are given by

$$\Delta = \frac{\mu_0 M_S H_{k,eff} V_{FL}}{2 k_B T} \quad (1)$$

$$I_c = \frac{\alpha e \gamma \mu_0 M_S H_{k,eff} V_{FL}}{\mu_B g_{STT}} \quad (2)$$

where $\mu_0$ is the vacuum permeability, $H_{k,eff}$ is the effective anisotropy field, $V_{FL}$ is the FL volume, $k_B$ is the Boltzmann constant, *T* is the FL temperature, $\alpha$ is the Gilbert damping factor, *e* is the electron charge, $\gamma$ is the gyromagnetic ratio, $\mu_B$ is the Bohr magneton, and $g_{STT}$ is the STT spin efficiency term depending on the *P* [13]. Owing to the decrease of $M_S$ and $K_i$ with increasing temperature (see Table 1) and according to (1)-(2), both $\Delta$ and $I_c$ expectedly tend to decrease as temperature increases [23], as shown in Fig. 2(b) and (c). Then, Fig. 2(d) shows the write error rate (WER) vs. bias voltage ($V_{MTJ}$) curves for 0→1 switching transition (i.e., from $R_H$ to $R_L$) at a pulse width ($t_{pulse}$) of 10 ns and different temperatures. According to the decrease of $R_H$, $\Delta$ and $I_c$ with increasing temperature, the bias voltage for a target WER ($10^{-7}$ in Fig. 2(d)) decreases as temperature increases. In turn, for a given bias voltage, the WER decreases with increasing temperature.

## 3. SIMPLY Logic Scheme

Fig. 3(a) illustrates the basic architecture of the SIMPLY logic. It consists of two MTJs (P and Q) and a load resistor $R_G$ as in the conventional IMPLY scheme [6–9], along with an additional comparator. A control logic block equipped with analog tri-state buffers as in the conventional IMPLY scheme drives the top electrodes of the MTJs with appropriate voltages depending on the operation to be performed and the output voltage of the comparator [9–12]. The inputs of the comparator are driven by the voltage $V_G$ developed across $R_G$ and a proper reference voltage $V_{REF}$ as provided by a reference circuit. Within this scheme, the initially stored states of P and Q are the inputs of the logic operation and the output Q' is determined by the state of Q after processing. As reported in the truth table of Fig. 3(a), proper operation requires that P keeps its state regardless of the input combination, whereas Q must switch from 0 to 1 only for the case P=Q=0. The latter can be effectively distinguished from the other input combinations by performing a preliminary READ operation through the comparator. This consists of applying a voltage pulse with amplitude $V_{READ}$ and width $t_{READ}$ on both MTJ devices and then comparing the voltage $V_G$ with a proper reference voltage $V_{REF}$. The proper detection of the input combination P=Q=0 from the output of the comparator thus allows performing the subsequent SET operation only in this case through the control logic. This is implemented by applying a voltage pulse with amplitude $V_{SET}$ and width $t_{SET}$ on Q, while keeping the driver of P in high impedance (HI-Z) state, as shown in Fig. 3(b). Instead, for the other input combinations the control logic forces the drivers of both MTJ devices to the HI-Z state, as shown in Fig. 3(c), to guarantee significant energy saving [9–12]. From Fig. 3(a), note that the scheme is also able to implement the FALSE operation on a single device by applying a negative $V_{RESET}$ pulse [9].

## 4. Simulation Results and Discussion

This section presents and discusses simulation results of our study on the STT-MTJ based SIMPLY architecture. First, we report results obtained for a given $R_G$ value and bias condition at room temperature (*T* = 300 K). Then, we analyze the circuit performance in terms of error rate and energy consumption at 300 K when varying the $R_G$ value and applied READ/SET voltages. Finally, our study is extended to evaluate the effect of temperature variations on circuit operation. All simulation results reported below were obtained by means of electrical simulations within Cadence Virtuoso environment. In particular, simulation data which include the effect of MTJ process variations were obtained from Monte Carlo simulations based on 1,000 runs. According to the scheme of Fig. 3(a), performed simulations included the two MTJ devices modeled through our Verilog-A compact model [13], the resistor $R_G$ and the comparator, which was designed in a 45-nm CMOS technology using the sense amplifier circuit shown in Fig. 3(a) [11] and whose contribution was considered only in terms of energy consumption [12]. The remaining circuit blocks were implemented by using a simplified approach. More specifically, we employed ideal voltage sources both for analog tri-state drivers providing $V_{READ}/V_{SET}$/HI-Z to the top electrode of the MTJ devices and the reference circuit providing the $V_{REF}$ voltage at the input of the comparator, whereas the control logic was functionally modeled by properly setting the output of MTJ drivers according to the operation to be executed.

Fig. 4(a)-(c) summarize the simulation results obtained with $R_G$ = 10 kΩ at room temperature for both READ and SET operations. Concerning the preliminary READ operation, its reliability is determined by two sources of error: (*i*) the read disturbance rate (RDR), i.e., the probability of unintentionally switching the stored data during READ operation [24], and (*ii*) the bit error rate (BER), i.e., the failure probability in distinguishing the input combination P=Q=0 from the others. In Fig. 4, we set $V_{READ}$ = 0.35 V and $t_{READ}$ = 10 ns, thus resulting into a RDR in the order of $10^{-9}$ [24, 25] for P=Q=0 (i.e., the worst case). Accordingly, Fig. 4(a)-(b) show the $V_G$ statistical distributions obtained for the different input combinations during the preliminary READ operation, while highlighting the corresponding estimated values for the read margin (RM), BER and $V_{REF}$. Such results were extracted from Monte Carlo simulations while considering the effect of MTJ process variations. The latter are modeled in our Verilog-A code by setting the variability (i.e., the ratio of the standard deviation to the mean value) of some parameters, whose variations are described through a Gaussian distribution. More precisely, we set a variability equal to 1% and 5% for the oxide thickness ($t_{OX}$) and the cross-section area, respectively [13, 26]. From Fig. 4(a), obtained results lead to a nominal RM (i.e., $RM_{nom}$ as defined by the difference between the mean values μ of $V_G$ distributions for the cases P≠Q and P=Q=0, respectively) equal to ~41 mV. The corresponding RM evaluated at the 3σ corner ($RM_{3σ}$) with σ being the standard deviation of $V_G$ distributions is 10.6 mV. From Fig. 4(b), the $V_{REF}$ is estimated as the voltage value (150.8 mV) that leads to the same BER ($2.6×10^{-5}$) for the input combinations P≠Q and P=Q=0. However, Fig. 4(b) also reveals that a small change in $V_{REF}$ can translate into a notable BER degradation. Taking this into account and in order to consider more effectively the effect of different RM values on the reliability of the READ operation (as in the analysis below), we evaluate the BER while assuming a change of ±5 mV in the $V_{REF}$. Such $V_{REF}$ variations result into a BER increase up to $1.7×10^{-3}$ and $7.8×10^{-4}$ for the cases P=Q=0 and P≠Q, respectively, as shown in Fig. 4(b). It is worth pointing out that, besides $V_G$ and $V_{REF}$ variations, imbalances in the comparator circuit branches represent another source of uncertainty for the READ operation [27], whose effect was neglected in our analysis. Such additional contribution typically translates into further BER degradation and can be effectively counteracted along with the effect of $V_{REF}$ variations by implementing boosted sensing schemes (i.e., by amplifying the detected voltage), as discussed and proposed in [27]. Fig. 4(c) shows data referred to the SET operation for the case P=Q=0 in terms of WER vs. $V_{SET}$ at $t_{SET}$ = 10 ns. According to Fig. 2(d), the WER decreases with increasing pulse amplitude. In Fig. 4(c), $V_{SET}$ is chosen to ensure a reasonably low WER = $10^{-7}$ [28], which also prevents further error degradation for the case P=Q=0. Table 2 then summarizes the results obtained in the above analysis in terms of both error rate and energy consumption. Here, energy data includes the additional contribution of the comparator. Data reported in Table 2 highlight that the overall error of the STT-MTJ based SIMPLY scheme is mainly determined by the BER referred to the preliminary READ operation. From Table 2, we obtained an average error (i.e., averaged across the four input combinations) equal to $8.2×10^{-4}$ with an average energy consumption of 160.1 fJ.

Fig. 5-7 show simulation results obtained at 300 K by varying the $R_G$ value (from 5 kΩ up to 30 kΩ) and applied voltages for READ and SET operations ($V_{READ}$ from 0.2 V up to 1 V and $V_{SET}$ from 0.5 V up to 1.3 V), while keeping $t_{READ}$ = $t_{SET}$ = 10 ns.

Fig. 5(a)-(c) show the color maps of the $RM_{nom}$, $RM_{3σ}$, and $V_{REF}$ referred to the preliminary READ operation. Observed trends are mainly explained by the fact that, for a given $R_G$ value, an increase in $V_{READ}$ results into higher current flowing through $R_G$ and hence higher μ and σ values related to $V_G$ distributions for the cases P≠Q and P=Q=0. From Fig. 5(a), the increased mean values of $V_G$ distributions with increasing $V_{READ}$ translates into an increase of the $RM_{nom}$, which is partly limited at very large $V_{READ}$ values by the *TMR* degradation owing to the increasing voltage drop across MTJ devices [13]. Moreover, Fig. 5(a) suggests that larger $RM_{nom}$ values are reachable for higher $R_G$. The increase of both μ and σ of $V_G$ distributions with increasing $V_{READ}$ leads to a maximum in the $RM_{3σ}$ for a given $R_G$ value, as shown in Fig. 5(b). This figure also shows that larger $R_G$ values are beneficial to achieve higher $RM_{3σ}$. Fig. 5(c) shows the $V_{REF}$ extracted from $V_G$ distributions at different $R_G$ and $V_{READ}$ values in the way described in Fig. 4(b), i.e., the voltage value that makes the BER for the input combinations P≠Q and P=Q=0 exactly the same. From Fig. 5(c), the $V_{REF}$ increases for larger $V_{READ}$ and $R_G$ values following the corresponding increase of the μ of $V_G$ distributions.

Fig. 6(a)-(c) show the color maps of error rates for both READ and SET operations, i.e., the average RDR, the average BER and the WER for the case P=Q=0. From Fig. 6(a), for a given $R_G$ value the average RDR increases with increasing $V_{READ}$ due to the corresponding increase in the current flowing through MTJ devices. Conversely, for a given $V_{READ}$ value the RDR is reduced at larger $R_G$ due to the lowering of the MTJ current. From Fig. 6(b), the average BER (again evaluated considering a change of ±5 mV with respect to $V_{REF}$ values reported in Fig. 5(c)) exhibits a roughly specular trend to that of the $RM_{3σ}$ shown in Fig. 5(b). More specifically, for a given $R_G$ value we can observe a minimum

BER, whereas larger $R_G$ values allow achieving lower BER, but at larger $V_{READ}$. For instance, the minimum BER for $R_G$ = 5 kΩ is $3.7 \times 10^{-3}$ at $V_{READ}$ = 0.35 V and decreases down to $1.8 \times 10^{-4}$ for $R_G$ = 30 kΩ at $V_{READ}$ = 0.6 V. From Fig. 6(c), contrary to the RDR, the WER decreases (increases) for larger $V_{SET}$ ($R_G$) values due to the increase (decrease) in the current flowing through the MTJ device. As a consequence, the $V_{SET}$ required to ensure the target WER of $10^{-7}$ increases with increasing $R_G$, i.e., from 0.67 V at $R_G$ = 5 kΩ up to 1.24 V at $R_G$ = 30 kΩ, thus resulting in an increase of the energy consumption for the SET operation.

Fig. 7(a)-(b) then show the color maps of the average error and average energy consumption when varying $R_G$ and $V_{READ}$ values. Here, for the different values of $R_G$ the $V_{SET}$ is considered fixed to the value needed to reach the WER = $10^{-7}$, as shown in Fig. 6(c). From the comparison between Fig. 6(a)-(c) and Fig. 7(a), we can again observe that the overall error is mainly determined by the BER related to the READ operation, except for very large $V_{READ}$ values where the RDR significantly degrades. Accordingly, Fig. 7(a) shows a minimum error at a specific $V_{READ}$ for a given $R_G$ value. In addition, as $R_G$ increases, such minimum error decreases and is achieved at larger $V_{READ}$ values. However, the decrease in the error at larger $R_G$ values corresponds to an increase in the energy consumption, as shown in Fig. 7(b). In particular, we can note that for $R_G$ = 30 kΩ a minimum error of $1.8 \times 10^{-4}$ is achieved at $V_{READ}$ = 0.6 V along with a corresponding energy consumption of 201.8 fJ. An energy saving of ~21% (159.4 fJ) is obtained at $R_G$ = 15 kΩ and $V_{READ}$ = 0.375 V, while keeping the error in the same order of magnitude ($4 \times 10^{-4}$). Such results thus highlight an existing tradeoff between error rate and energy consumption, which can be effectively managed by properly setting the load resistor and applied voltages.

Finally, Fig. 8(a)-(f) summarize the simulation results obtained under temperature variations in the range from 250 K up to 350 K. Data refer to the case with $R_G$ = 15 kΩ, $t_{READ}$ = $t_{SET}$ = 10 ns, $V_{READ}$ = 0.375 V, and $V_{SET}$ = 0.89 V. Fig. 8(a) shows the average RDR, which increases with increasing temperature by a factor of ~$10^8$ in the considered range. This is ascribed to the increase in the current flowing through MTJ devices owing to the decrease of $R_H$ (see Fig. 2(a)), as well as to the decrease of both Δ and $I_c$ as shown in Fig. 2(b)-(c). Conversely, the WER referred to the case P=Q=0 decreases as temperature increases by a factor of 144 when the temperature goes from 250 K up to 350 K, as shown in Fig. 8(b). Fig. 8(c) shows the temperature dependence of the $RM_{nom}$ and $RM_{3σ}$, both decreasing with increasing temperature (-15% and -40%, respectively, in the considered range) as mainly given by the decrease in the *TMR* ratio (see Fig. 2(a)). Then, Fig. 8(d) shows the trend of the $V_{REF}$ with temperature as determined by the temperature effect on the MTJ properties. Reported data suggest that the $V_{REF}$ should exhibit a proportional to absolute temperature (PTAT) behavior to keep low the BER of the READ operation and hence the overall error across temperatures. This can be appreciated in Fig. 8(e), which illustrates the average error evaluated while considering again a change of ±5 mV in the $V_{REF}$ and referred to two cases: (*i*) constant $V_{REF}$ with temperature to the value extracted at 300 K in Fig. 8(d), i.e., assuming a voltage reference circuit that provides an ideally stable $V_{REF}$ (±5 mV) across temperatures, and (*ii*) $V_{REF}$ with PTAT trend as in Fig. 8(d), i.e., assuming a voltage reference circuit that provides a PTAT $V_{REF}$ (±5 mV) based on the predicted behavior of the memory cell. From Fig. 8(e), tracking the temperature dependence of the MTJ properties by a PTAT $V_{REF}$ allows keeping the average error below $1.1 \times 10^{-3}$ in the considered temperature range with an improvement of ×5.4 and ×4.4 at 250 K and 350 K, respectively, compared to the case with constant $V_{REF}$. Nonetheless, the estimated average error increases with increasing temperature even with the use of a PTAT $V_{REF}$, according to reduced read margins shown in Fig. 8(c). Finally, Fig. 8(f) illustrates the average energy consumption (including the contribution of the comparator) under temperature variations, which slightly decreases by -8% when the temperature goes from 250 K up to 350 K. It is worth pointing out that energy data of Fig. 8(f) refer to both analyzed cases (i.e., constant or PTAT $V_{REF}$ across temperatures), considering that the difference between $V_{REF}$ values provided in the two cases at a given temperature (see Fig. 8(d)) does not imply any significant change in the energy consumption of the comparator.

## 5. Conclusion

In this work, we investigated the SIMPLY logic architecture relying on nanoscale STT-MTJ devices. In particular, we evaluated the circuit performance by varying the load resistor and applied voltages to implement both READ and SET operations, while also analyzing the circuit operation as temperature changes. Our study was performed by using a macrospin-based Verilog-A compact model to describe the behavior and characteristics of a perpendicular STT-MTJ device with circular geometry and diameter of 30 nm.

As main result of our analysis, an existing tradeoff between error rate and energy consumption is demonstrated. Such tradeoff can be effectively managed by properly setting the load resistor and applied voltages. To give a reference, the average error obtained for $R_G$ = 30 kΩ and $V_{READ}$ = 0.6 V is $1.8 \times 10^{-4}$ with a corresponding average energy of 201.8 fJ, whereas the use of $R_G$ = 15 kΩ along with $V_{READ}$ = 0.375 V results into an average error of $4 \times 10^{-4}$ and an average energy of 159.4 fJ. Obtained results also prove a degradation of the error rate under temperature variations. However, we demonstrated that the use of a PTAT reference voltage at the input of the comparator allows tracking the temperature dependence of the MTJ properties, thus mitigating the effect of temperature variations on the circuit reliability.

## Acknowledgment

This work was partially supported by the project PRIN 2020LWPKH7 funded by the Italian Ministry of University and Research.

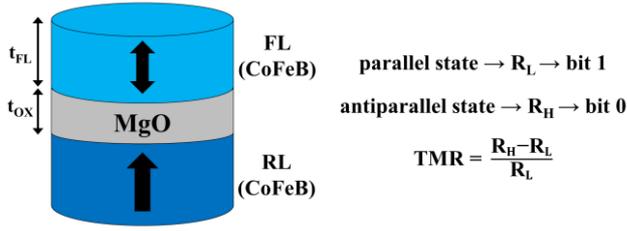

**Fig. 1.** STT-MTJ sketch and resistive states.

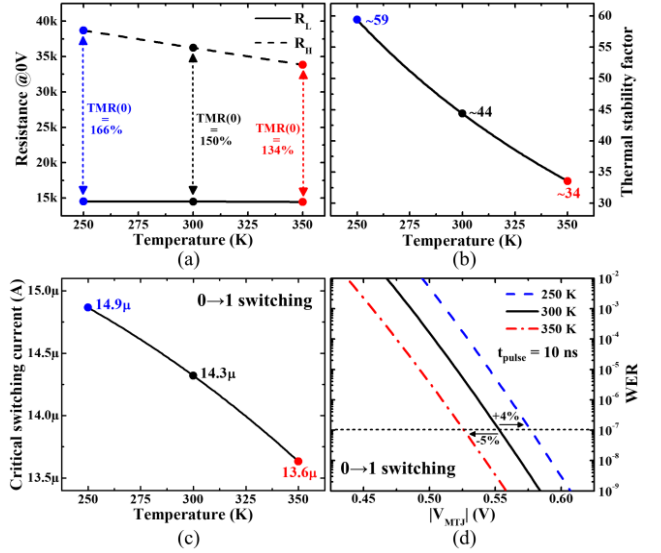

**Fig. 2.** Temperature-dependent STT-MTJ characteristics: (a) resistance in low ($R_L$) and high ($R_H$) states at zero bias voltage with the detail of corresponding tunnel magnetoresistance (*TMR*) ratio, (b) thermal stability factor (Δ), (c) critical switching current ($I_c$) for 0→1 switching (i.e., from antiparallel to parallel state), (d) write error rate (WER) vs. bias voltage ($V_{MTJ}$) for 0→1 switching at pulse width $t_{pulse}$ = 10 ns.

TABLE 1
STT-MTJ Parameters

| Parameter | Description (unit) | Value (250 K, 300 K, 350 K) |
|---|---|---|
| $d$ | Diameter (nm) | 30 |
| $t_{FL}$ | FL thickness (nm) | 1.15 |
| $t_{OX}$ | Oxide barrier thickness (nm) | 0.85 |
| RA | Resistance-area product in parallel state ($\Omega \cdot \mu m^2$) | 10 [14] |
| P | Spin polarization factor | (0.68, 0.66 [15], 0.64) |
| $V_H$ | Bias voltage for $TMR = TMR(0)/2$ (V) | 0.5 |
| $M_S$ | Saturation magnetization (T) | (1.64, 1.58 [16], 1.51) |
| α | Gilbert damping factor | 0.03 |
| $K_i$ | Interfacial perpendicular anisotropy constant (mJ/m$^2$) | (1.41, 1.3 [16], 1.18) |

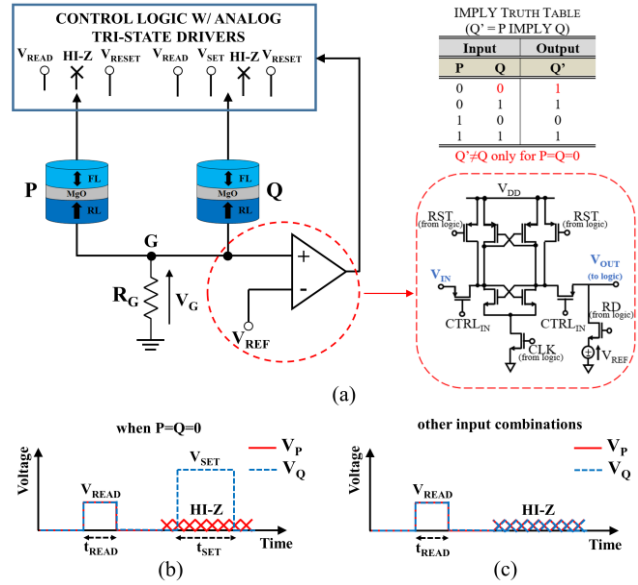

**Fig. 3.** (a) Scheme of the STT-MTJ based SIMPLY scheme with the detail of comparator and truth table of the IMPLY logic operation; time diagram of applied voltage pulses (b) when the comparator detects the input condition P=Q=0 and (c) in all other cases. Taken from [12].

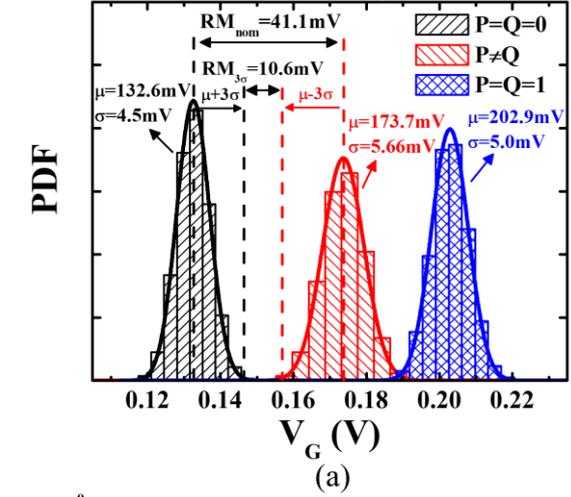

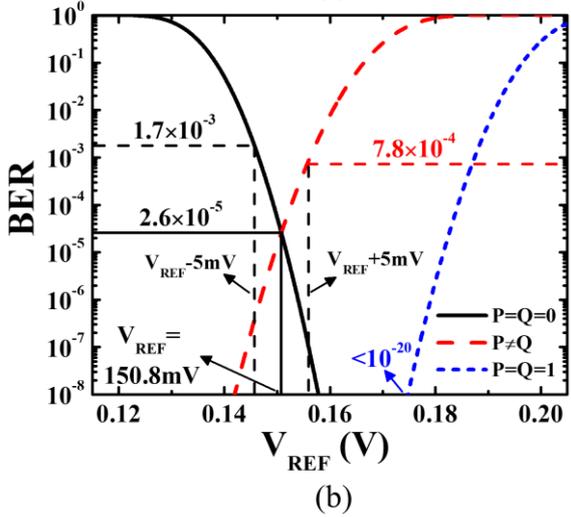

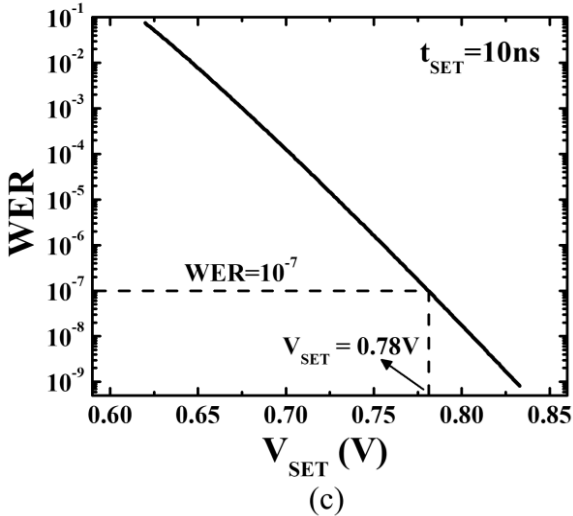

**Fig. 4.** Simulation results with $R_G = 10$ k$\Omega$ at $T = 300$ K for READ ($V_{READ} = 0.35$ V and $t_{READ} = 10$ ns) and SET ($t_{SET} = 10$ ns) operations. (a) $V_G$ statistical distributions for the different input combinations under MTJ process variations related to the preliminary READ operation along with (b) the estimation of the bit error rate (BER) and reference voltage $V_{REF}$. (c) WER vs. $V_{SET}$ at the nominal corner (i.e., no process variations) for the case P=Q=0.

TABLE 2
Summary results from Fig. 4 ($T = 300$ K, $R_G = 10$ k$\Omega$, $V_{READ} = 0.35$ V, $t_{READ} = 10$ ns, $V_{REF} = 150.8$ mV, $V_{SET} = 0.78$ V, $t_{SET} = 10$ ns)

| P | Q | RDR | BER* | WER | Error* | Energy** (J) |
|---|---|---|---|---|---|---|
| 0 | 0 | $8.9 \times 10^{-10}$ | $1.7 \times 10^{-3}$ | $10^{-7}$ | $1.7 \times 10^{-3}$ | 318.2f |
| 0 | 1 | $5.5 \times 10^{-12}$ | $7.8 \times 10^{-4}$ | --- | $7.8 \times 10^{-4}$ | 104.2f |
| 1 | 0 | $5.5 \times 10^{-12}$ | $7.8 \times 10^{-4}$ | --- | $7.8 \times 10^{-4}$ | 104.2f |
| 1 | 1 | 0 | $<10^{-20}$ | --- | $<10^{-20}$ | 113.9f |

* at $V_{REF} \pm 5$mV
** including the contribution of the comparator implemented in a 45-nm CMOS technology [11]

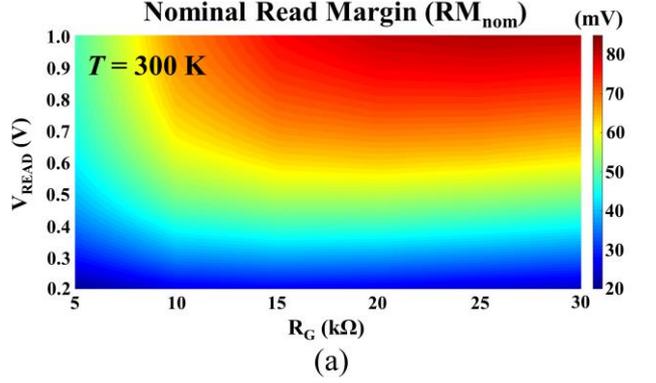

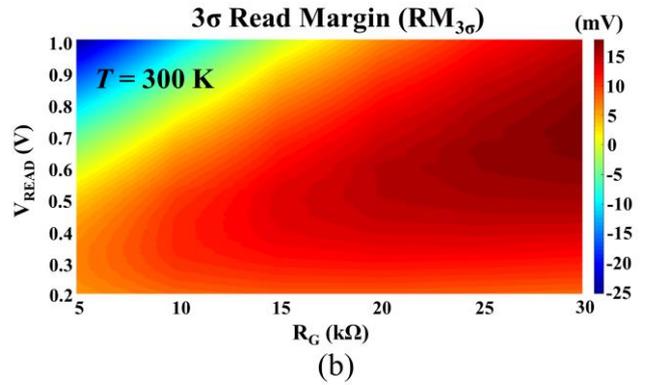

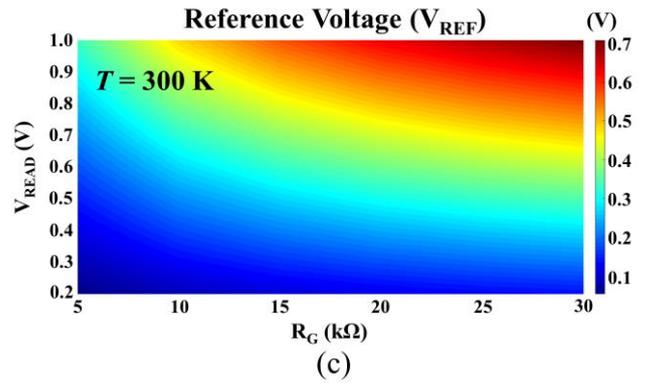

**Fig. 5.** Simulation results at $T = 300$ K for the READ operation by varying $R_G$ and $V_{READ}$: (a) nominal read margin ($RM_{nom}$), (b) read margin at the $3\sigma$ corner ($RM_{3\sigma}$), and (c) reference voltage ($V_{REF}$).

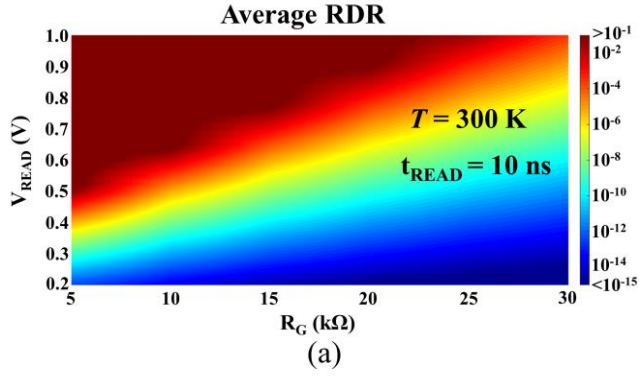

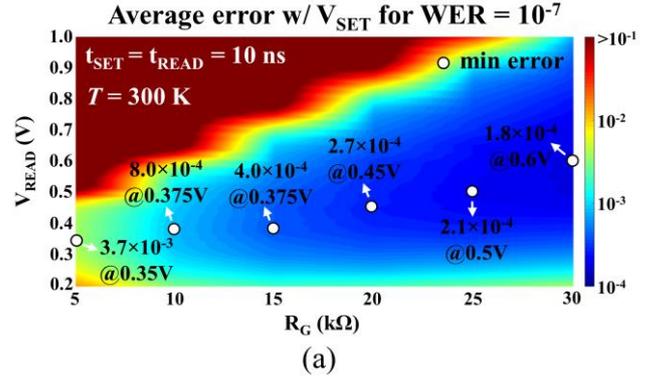

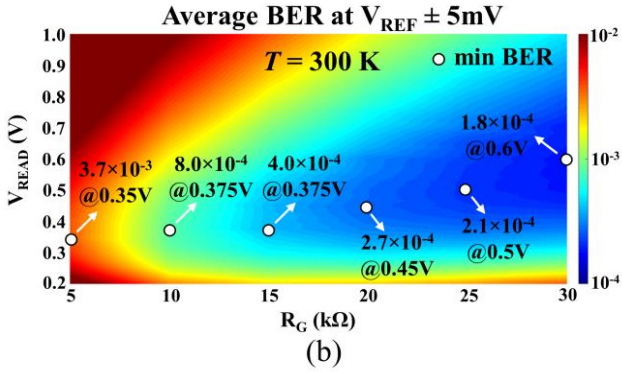

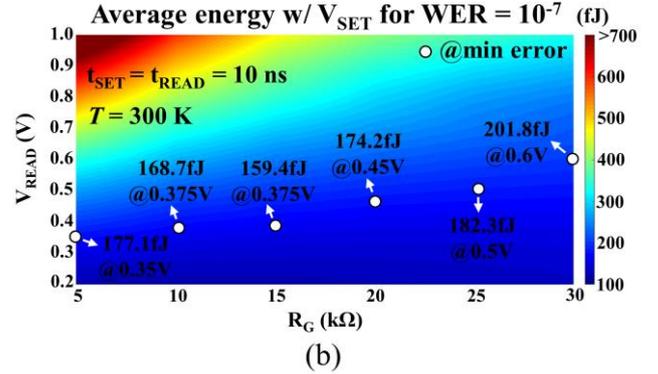

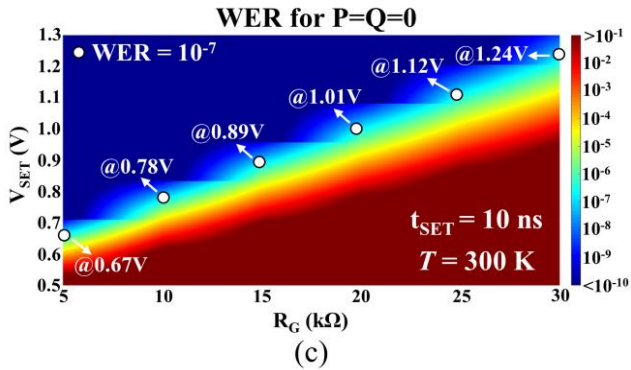

**Fig. 6.** Simulation results at $T$ = 300 K for READ ($t_{READ}$ = 10 ns) and SET ($t_{SET}$ = 10 ns) operations by varying $R_G$, $V_{READ}$, and $V_{SET}$: (a) average read disturbance rate (RDR), (b) average bit error rate (BER) evaluated at $V_{REF}$ ±5 mV, (c) write error rate (WER) for the case P=Q=0.

**Fig. 7.** Simulation results at $T$ = 300 K in terms of (a) average error and (b) average energy consumption (including the contribution of the comparator) by varying $R_G$ and $V_{READ}$ with $t_{READ}$ = $t_{SET}$ = 10 ns and $V_{SET}$ to ensure a WER = $10^{-7}$.

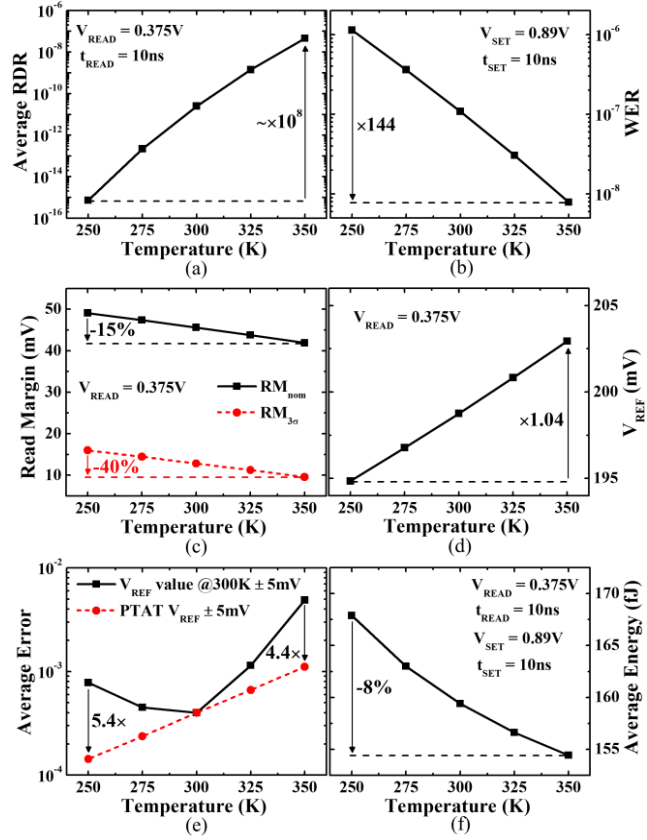

**Fig. 8.** Simulation results under temperature variations for $R_G = 15$ kΩ, $t_{READ} = t_{SET} = 10$ ns, $V_{READ} = 0.375$ V, $V_{SET} = 0.89$ V. (a) Average read disturbance rate (RDR). (b) Write error rate (WER) for the case P=Q=0. (c) Nominal read margin ($RM_{nom}$) and read margin at the 3σ corner ($RM_{3\sigma}$). (d) Reference voltage ($V_{REF}$). (e) Average error with constant $V_{REF}$ with temperature to the value extracted at 300 K or with proportional to absolute temperature (PTAT) $V_{REF}$ as reported in (d), while considering a change of ±5 mV in the $V_{REF}$. (f) Average energy consumption (including the contribution of the comparator).